\documentclass[twocolumn]{aastex631}

\usepackage{amsmath}

\begin{document}

\title{Measuring Milky Way Mass with RR Lyraes and Blue Horizontal Branch Stars}

\author[0009-0006-1196-574X]{Xiangyu Li}
\affiliation{National Astronomical Observatories, Chinese Academy of Sciences, Beijing 100101, China}
\affiliation{School of Astronomy and Space Science, University of Chinese Academy of Sciences, Beijing 100049, China}

\author[0000-0003-3250-2876]{Yang Huang}
\affiliation{School of Astronomy and Space Science, University of Chinese Academy of Sciences, Beijing 100049, China}
\affiliation{National Astronomical Observatories, Chinese Academy of Sciences, Beijing 100101, China}

\author{Jifeng Liu}
\affiliation{National Astronomical Observatories, Chinese Academy of Sciences, Beijing 100101, China}
\affiliation{School of Astronomy and Space Science, University of Chinese Academy of Sciences, Beijing 100049, China}

\author[0000-0003-2086-0684]{Xinyi Li}
\affiliation{College of Physics and Electronic Engineering, Qilu Normal University, Jinan 250200, People’s Republic of China}

\author{Qikang Feng}
\affiliation{Department of Astronomy, School of Physics, Peking University, Beijing 100871, China}
\affiliation{Kavli Institute for Astronomy and Astrophysics, Peking University, Beijing 100871, China}

\author[0009-0002-4542-8046]{Haozhu Fu}
\affiliation{Department of Astronomy, School of Physics, Peking University, Beijing 100871, China}
\affiliation{Kavli Institute for Astronomy and Astrophysics, Peking University, Beijing 100871, China}

\author[0000-0002-9796-1507]{Shenglan Sun}
\affiliation{Department of Astronomy, School of Physics, Peking University, Beijing 100871, China}
\affiliation{Kavli Institute for Astronomy and Astrophysics, Peking University, Beijing 100871, China}

\author{Jiahui Wang}
\affiliation{National Astronomical Observatories, Chinese Academy of Sciences, Beijing 100101, China}
\affiliation{School of Astronomy and Space Science, University of Chinese Academy of Sciences, Beijing 100049, China}

\correspondingauthor{Yang Huang (huangyang@ucas.ac.cn)}

\begin{abstract}

Measuring the mass of the Milky Way is essential for understanding galaxy formation and evolution, characterizing the Galactic potential and assembly history, and probing the nature of dark matter. We derive the enclosed mass profile of the Milky Way out to Galactocentric distances of $\sim80\,\mathrm{kpc}$ using two halo tracers with full 3D kinematics: RR Lyraes (RRLs) and blue horizontal-branch stars (BHBs). Our sample consists of 7{,}550 RRLs and 8{,}230 BHBs constructed from the Large Sky Area Multi-Object Fiber Spectroscopic Telescope (LAMOST), the Sloan Digital Sky Survey (SDSS), and the Dark Energy Spectroscopic Instrument (DESI), spanning Galactocentric distances from 10 to $100\,\mathrm{kpc}$. 
We estimate the enclosed mass using two independent methods: the spherical Jeans equation and the tracer mass estimator (TME). The two tracers yield consistent results, with $M(<81.7\,\mathrm{kpc}) = (4.76 \pm 0.23)\times10^{11}\,\mathrm{M}_\odot$ from RRLs and $M(<82.8\,\mathrm{kpc}) = (4.66 \pm 0.35)\times10^{11}\,\mathrm{M}_\odot$ from BHBs.
Adopting an NFW profile to extrapolate the enclosed mass beyond the outermost radius probed by our tracers, we infer virial masses of $M_{200} = 8.12^{+0.72}_{-0.65}\times10^{11}\,\mathrm{M}_\odot$ from RRLs and $M_{200} = 8.13^{+0.78}_{-0.67}\times10^{11}\,\mathrm{M}_\odot$ from BHBs.
We further validate our methodology using mock RRL and BHB samples constructed from two Auriga halos, finding that the inferred masses are underestimated by $15.5\%\pm2.8\%$ for RRLs and $11.9\%\pm3.4\%$ for BHBs, respectively. After correcting for these systematic errors, the Milky Way virial masses are $M_{200} = 9.79^{+0.93}_{-0.84}\times10^{11}\,\mathrm{M}_\odot$ from RRLs and $M_{200} = 9.24^{+0.95}_{-0.84}\times10^{11}\,\mathrm{M}_\odot$ from BHBs.

\end{abstract}

\keywords{stars: RRL - stars: BHB - Galaxy: halo - Galaxy: mass - Galaxy: kinematics and dynamics}

\section{Introduction \label{sec:1}}

\begin{figure*}
\centering
\plotone{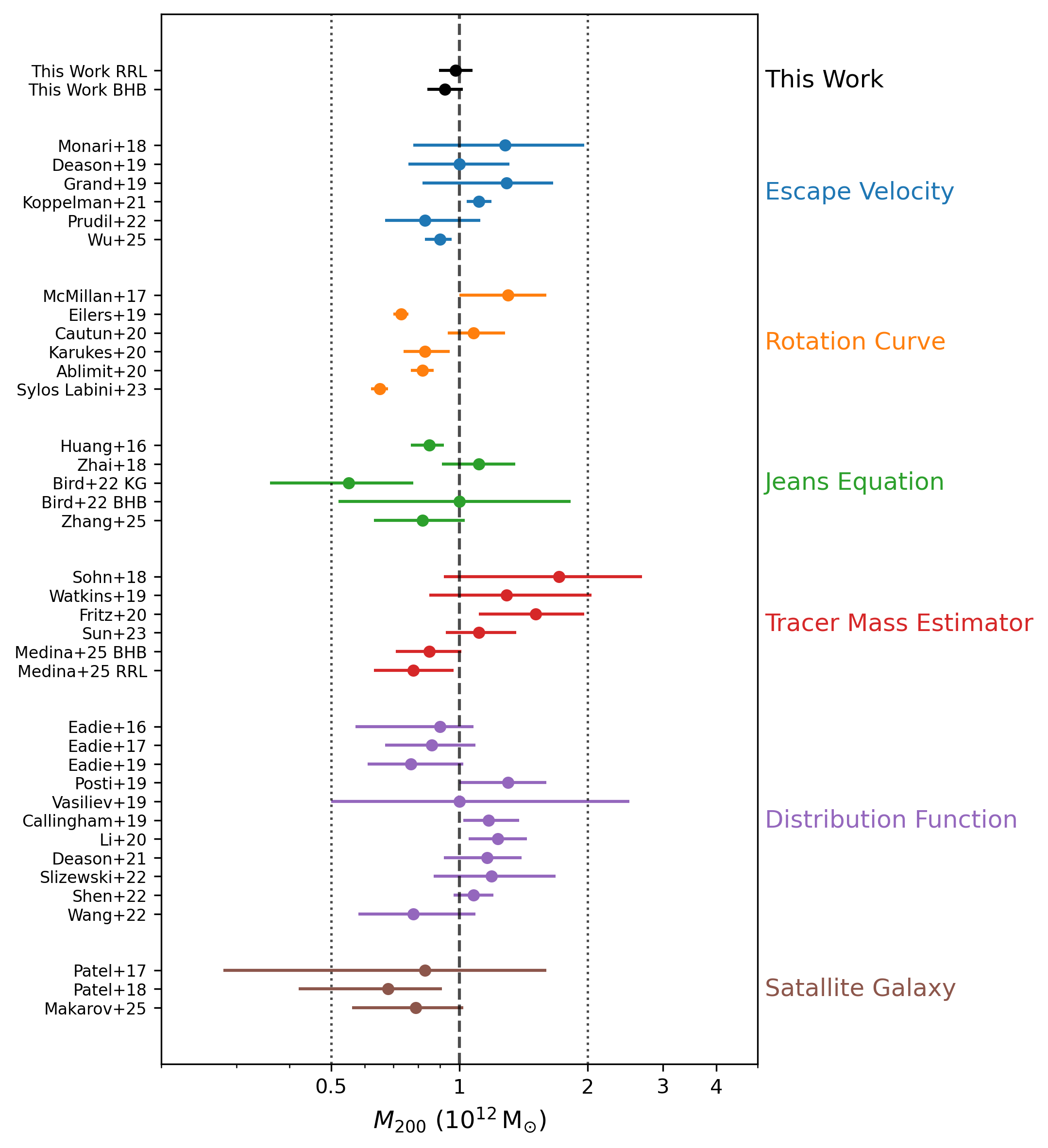}
\caption{Comparison of our virial mass estimates, after correcting for systematic biases identified in the simulated data, with values from the literature. The color of each data point represents the method employed in the corresponding work. The literature results are adopted from \citet{Wang2020, Monari2018, Deason2019, Grand2019, Koppelman2021, Prudil2022, Wu2025, McMillan2017, Eilers2019, Cautun2020, Karukes2020, Ablimit2020, Labini2023, Huang2016, Zhai2018, Bird, Zhang2025, Sohn2018, Watkins2019, Fritz2020, Sun2023, DESIMass, Eadie2016, Eadie2017, Eadie2019, Posti2019, Vasiliev2019, Callingham2019, Li2020, Deason2021, Slizewski2022, Shen2022, Wang2022, Patel2017, Patel2018, Makarov2025}.}
\label{fig:histo}
\end{figure*}

As our home galaxy, the Milky Way offers a unique opportunity to study hundreds of billions of stars in unprecedented detail within a multi-dimensional phase space. The total mass of the Milky Way is a fundamental parameter to place our Galaxy in a broader cosmological context, and is essential for understanding galaxy formation and evolution, testing the standard cosmological model, and probing the nature of dark matter \citep{ARAA}. 

The mass of the Milky Way is closely connected to a range of questions concerning the small-scale structure of the $\Lambda$CDM cosmological model. For example, the severity of the Too Big to Fail problem, which concerns the apparent overabundance of massive subhalos relative to the observed satellite population, depends sensitively on the assumed Milky Way halo mass \citep{TBTF}. Adopting a halo mass of $\sim2\times10^{12}\,\mathrm{M}\odot$ yields a less than $5\%$ probability of observing no more than three massive satellites, whereas a lower mass of $\sim1\times10^{12}\,\mathrm{M}\odot$ increases this probability to $\sim50\%$, substantially reducing the tension \citep{TBTF2}.

Despite its importance, estimates of the total mass of the Milky Way over the past two decades have exhibited discrepancies as large as $400\%$, as illustrated in Figure~\ref{fig:histo} (see a similar plot in recent review by \citealt{Wang2020}). A wide range of dynamical approaches and tracer populations have been employed to constrain the Milky Way mass. These include measurements based on the escape velocity of high-velocity stars \citep{Monari2018,Deason2019,Grand2019,Koppelman2021,Prudil2022,Wu2025}, the circular velocity of disk stars \citep{McMillan2017,Eilers2019,Cautun2020,Karukes2020,Ablimit2020,Labini2023,Huang2016,Zhai2018,Bird,Zhang2025}, and the dynamics of globular clusters, satellite galaxies, stellar streams, and halo stars \citep{Sohn2018,Watkins2019,Fritz2020,Sun2023,DESIMass,Eadie2016,Eadie2017,Eadie2019,Posti2019,Vasiliev2019,Callingham2019,Li2020,Deason2021,Slizewski2022,Shen2022,Wang2022,Patel2017,Patel2018,Makarov2025}. Among these, the Jeans equation, derived from the steady-state collisionless Boltzmann equation under the assumption of dynamical equilibrium, provide a widely used framework for relating the spatial distribution and kinematics of tracer populations to the underlying gravitational potential \citep[e.g.][]{Xue2008,Bird}. Despite the diversity of available methods and data sets, published measurements continue to span a broad range, indicating that the Milky Way mass remains far from being precisely established.

The origin of this large scatter can be understood in terms of two different radial regimes. Within $\sim80\,\mathrm{kpc}$, increasingly large samples of stellar tracers are now available, but their numbers remain limited, so the inferred mass is still affected primarily by statistical uncertainties in the tracer kinematics, including velocity measurement errors and the finite sampling of the underlying velocity distribution. At larger radii, the number of available tracers drops sharply, providing little direct constraint on the outer mass profile. Most estimates of the total or virial mass therefore require extrapolation beyond the radial range directly probed by the tracers, typically under assumptions about the dark matter halo profile, such as NFW profile. Differences in the adopted halo profiles and their parameters can introduce substantial systematic uncertainties in this extrapolation and may contribute significantly to the scatter among existing measurements. Improving the Milky Way mass estimate thus requires not only larger tracer samples to reduce statistical uncertainties in the directly probed region, but also tracers extending to larger Galactocentric radii to reduce the reliance on profile-dependent extrapolation.

The rapid growth of spectroscopic and astrometric surveys now provides an opportunity to address these limitations. Large spectroscopic surveys, including LAMOST \citep{LAMOST1,LAMOST2}, SDSS \citep{SDSS}, SEGUE \citep{SEGUE}, APOGEE \citep{APOGEE}, RAVE \citep{RAVE}, GALAH \citep{GALAH}, and DESI \citep{DESI}, together with the astrometric survey \textit{Gaia} \citep{Gaia}, have provided radial velocities, proper motions, distances, and chemical abundances for vast numbers of Galactic stars. In particular, the combination of spectroscopic radial velocities with precise \textit{Gaia} proper motions enables dynamical studies based on full three-dimensional velocities, substantially reducing the reliance on statistically inferred transverse motions. The resulting large and increasingly homogeneous samples of halo stars provide an opportunity to improve the statistical precision of the Milky Way mass profile and to extend direct dynamical constraints to larger Galactocentric radii. Among halo tracers, blue horizontal-branch (BHB) stars and RR Lyrae (RRL) stars are particularly well suited to this purpose because their well-calibrated luminosities enable reliable distance measurements over large Galactocentric radii, while their large numbers make them powerful probes of the outer halo.

In this work, we combine BHB and RRL samples from LAMOST, SDSS, and DESI with \textit{Gaia} astrometry to measure the enclosed mass profile of the Milky Way out to $\sim80\,\mathrm{kpc}$. We apply the spherical Jeans equation and the tracer mass estimator (TME) of \citet{Evans2011} as two independent approaches, and use mock BHB and RRL catalogs from the Auriga simulations to assess the systematic uncertainties of our analysis.

The structure of this paper is as follows. In Section \ref{sec:2}, we describe the construction of the BHB and RRL catalogs, the determination of their stellar density profiles, and the binning schemes adopted in our analysis. Section \ref{sec:3} presents the methods used to derive the enclosed mass profile and virial mass of the Milky Way. In Section \ref{sec:4}, we present our results and validate our methodology using mock BHB and RRL catalogs from the Auriga simulations. Finally, Section \ref{sec:5} discusses the sources of uncertainty, compares our results with previous measurements in the literature, and summarizes our conclusions.

\section{Data \label{sec:2}}

\subsection{Sample Selection \label{sec:2.1}}

\begin{figure*}
\centering
\plottwo{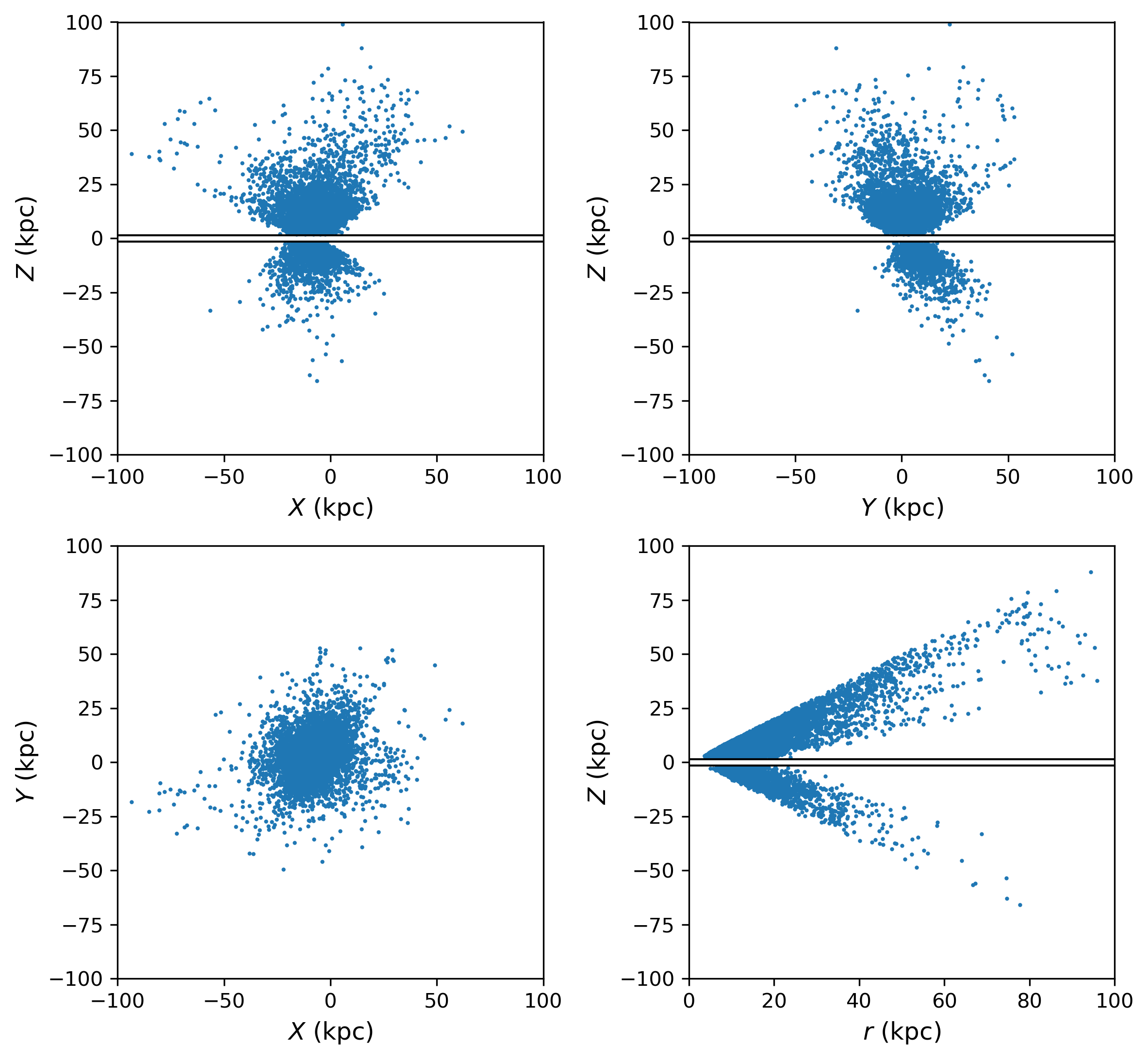}{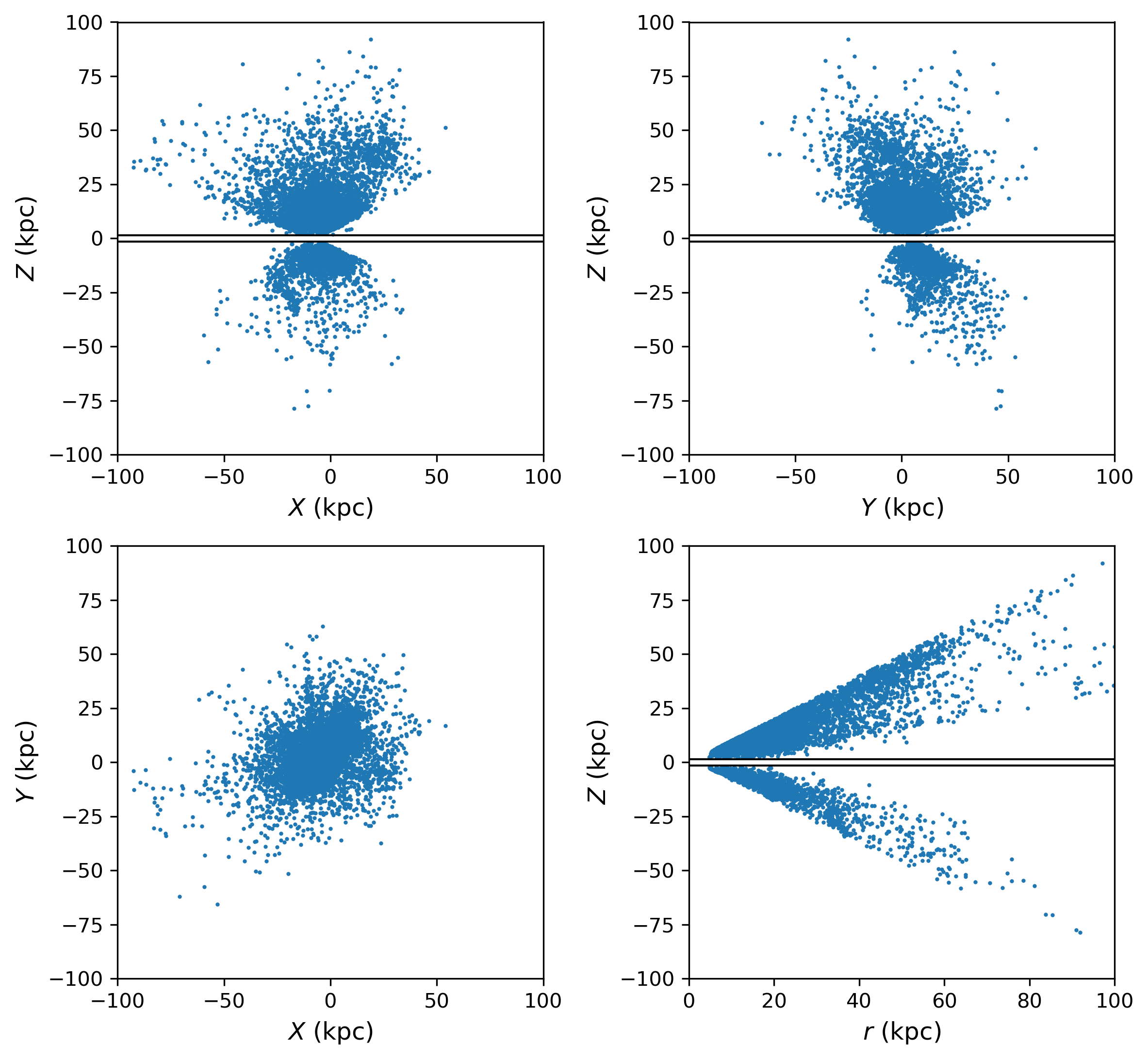}
\caption{Spatial distribution of the RRL (left) and BHB (right) samples (after substructure subtractions) in the Galactocentric Cartesian Coordinates. The black lines represent the boundary between the halo and the disk at $|Z|=2$\,kpc. $r$ is the Galactocentric distance defined by $r = \sqrt{X^2 + Y^2 + Z^2}$.}
\label{fig:xyz}
\end{figure*}

In this work, we employ two samples of stars as tracers of the Galactic halo: RRL and BHB stars. Both types of stars serve as standard candles, enabling accurate distance measurements even in large Galactic radius. 

We adopted the RRL sample from \citet{RRL1, Li2023} and \citet{He2025}, and the BHB sample from \citet{Xue2011} and \citet{DESIBHB}. These samples consist of line-of-sight velocities and metallicities obtained with LAMOST \citep{LAMOST1, LAMOST2}, SDSS \citep{SDSS, SEGUE, APOGEE} and DESI \citep{DESI}, distance estimates using the method of \citet{Li2023, He2025} and \citet{DESIBHB}, and proper motions from \textit{Gaia} DR3 \citep{Gaia} astrometric measurements.

The Galactocentric Cartesian and Spherical Coordinates both follow the conventions in \citet{Astropy}. We adopt the distance from the Sun to the Galactic center $R_\odot = 8.122\pm0.031$\,kpc \citep{X}, the vertical distance to the Galactic plane $Z_\odot = 20.8$\,pc \citep{Z}, and the Galactocentric velocity of the Sun $V_{R,\odot} = 11.1$\,km\,s$^{-1}$, $V_{\phi,\odot} = 245.6$\,km\,s$^{-1}$, and $V_{Z,\odot} = 7.8$\,km\,s$^{-1}$ \citep{UVW3, UVW4, Zhou2023}.

Both the Jeans equation and TME assume that the population tracing the Galactic potential is a relaxed and virialized system. Thus, we use the smooth, diffuse halo samples from which stellar streams and substructures have been removed using the method of \citet{Sun2025}, which identifies substructures through clustering in integral-of-motion space. The galactic disk has also been removed by limiting $|Z|>2$\,kpc.
The final sample consists of 7{,}550 RRLs and 8{,}230 BHBs covering Galactocentric distances out to 100\,kpc. Figure~\ref{fig:xyz} show the distribution of the two samples in the Galactocentric Cartesian Coordinates. 

\subsection{Tracer Density Profile \label{sec:2.2}}

The tracer density profile, $\rho(r)$, is an important ingredient in the mass determination. In both methods, we model the tracer density profile as a power law, with power-law index $\alpha$.

Previous studies have commonly adopted either a single or a broken power-law profile to describe the spatial distribution of halo tracers. To account for the perturbing effects of substructures, particularly the Gaia–Enceladus–Sausage (GES) and the Sagittarius (Sgr) dwarf galaxy, we adopt a broken power-law model with a fixed break radius of $r_\mathrm{break}=20\,\mathrm{kpc}$ for both tracer populations. By taking the mean values of the density-profile slopes reported in multiple literature studies \citep{RD1, RD2, RD3, RD4, RD5, BD1, BD2, BD3, BD4}, we obtain $\alpha_\mathrm{in}=-2.8$ and $\alpha_\mathrm{out}=-4.37$ for RRLs, and $\alpha_\mathrm{in}=-3.5$ and $\alpha_\mathrm{out}=-4.45$ for BHBs.

\subsection{Radial Binning \label{sec:2.3}}

To calculate the enclosed mass profile using the Jeans equation and the TME, we divide the stellar samples into radial bins according to their Galactocentric distances. We choose the bin boundaries by considering both the distance uncertainties and the number of stars available in each radial range. For both tracer populations, we adopt bin boundaries at $r=$ 10, 15, 20, 25, 35, 50, 70, and 100\,kpc. This binning scheme provides $500\sim1{,}500$ stars per bin within 50\,kpc and more than 100 stars per bin beyond 50\,kpc. For each bin, we use the median Galactocentric distance of the stars in that bin as the representative radius for the mass calculation and plotting.

\section{Methods \label{sec:3}}

\begin{figure}
\centering
\plotone{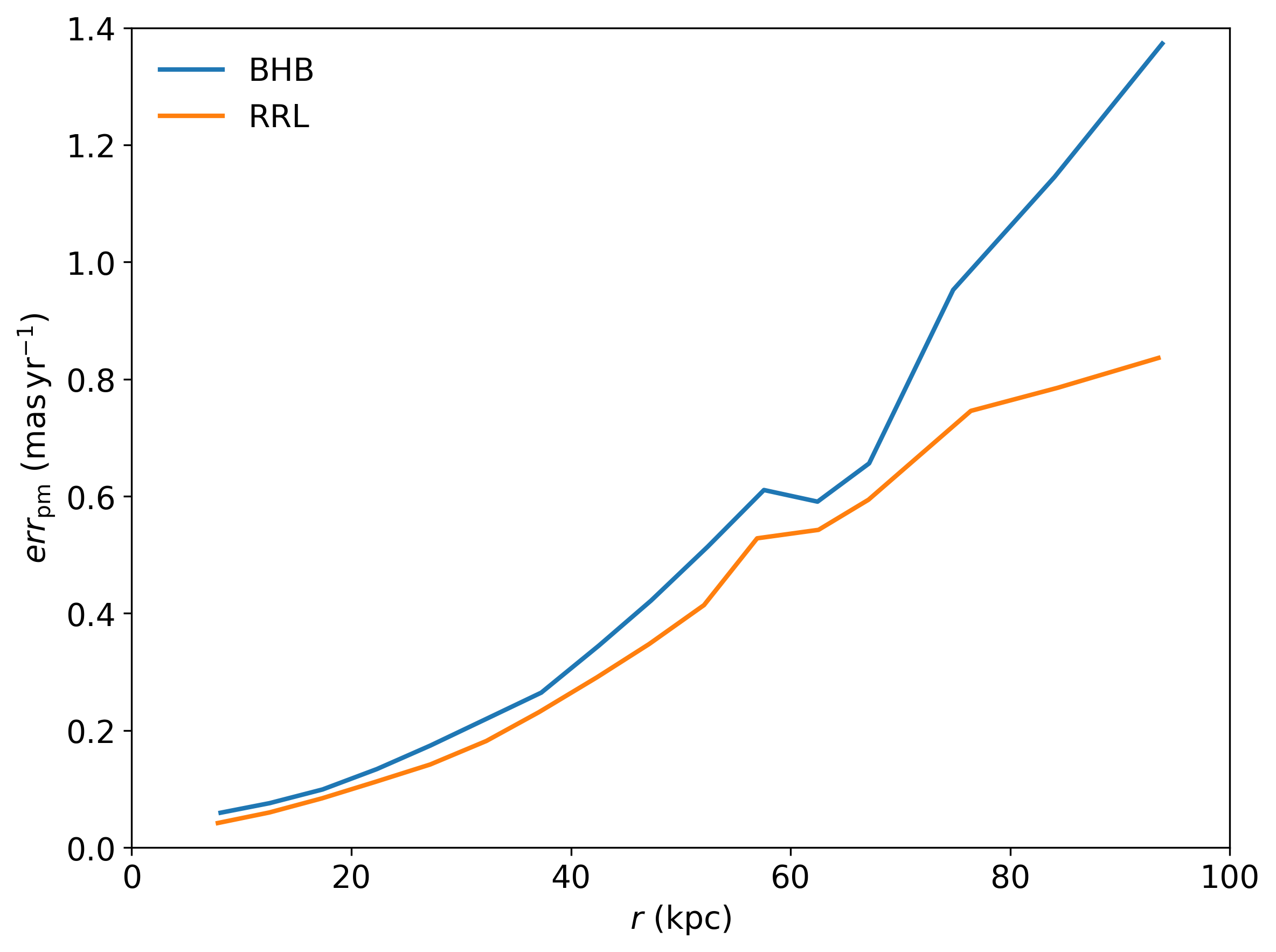}
\caption{Proper-motion uncertainty as a function of Galactocentric distance.}
\label{fig:pm}
\end{figure}

We apply two methods to our data to recover the Milky Way’s enclosed mass profile: the spherical Jeans equation and the tracer mass estimator (TME). Both methods require full 3D distances and velocities. We also extrapolate our enclosed mass profile to estimate the virial mass and radius of the Milky Way, assuming an NFW dark matter profile. In this section, we describe the details of each method and the associated uncertainties propagated from the observed data.

\begin{figure*}
\centering
\plottwo{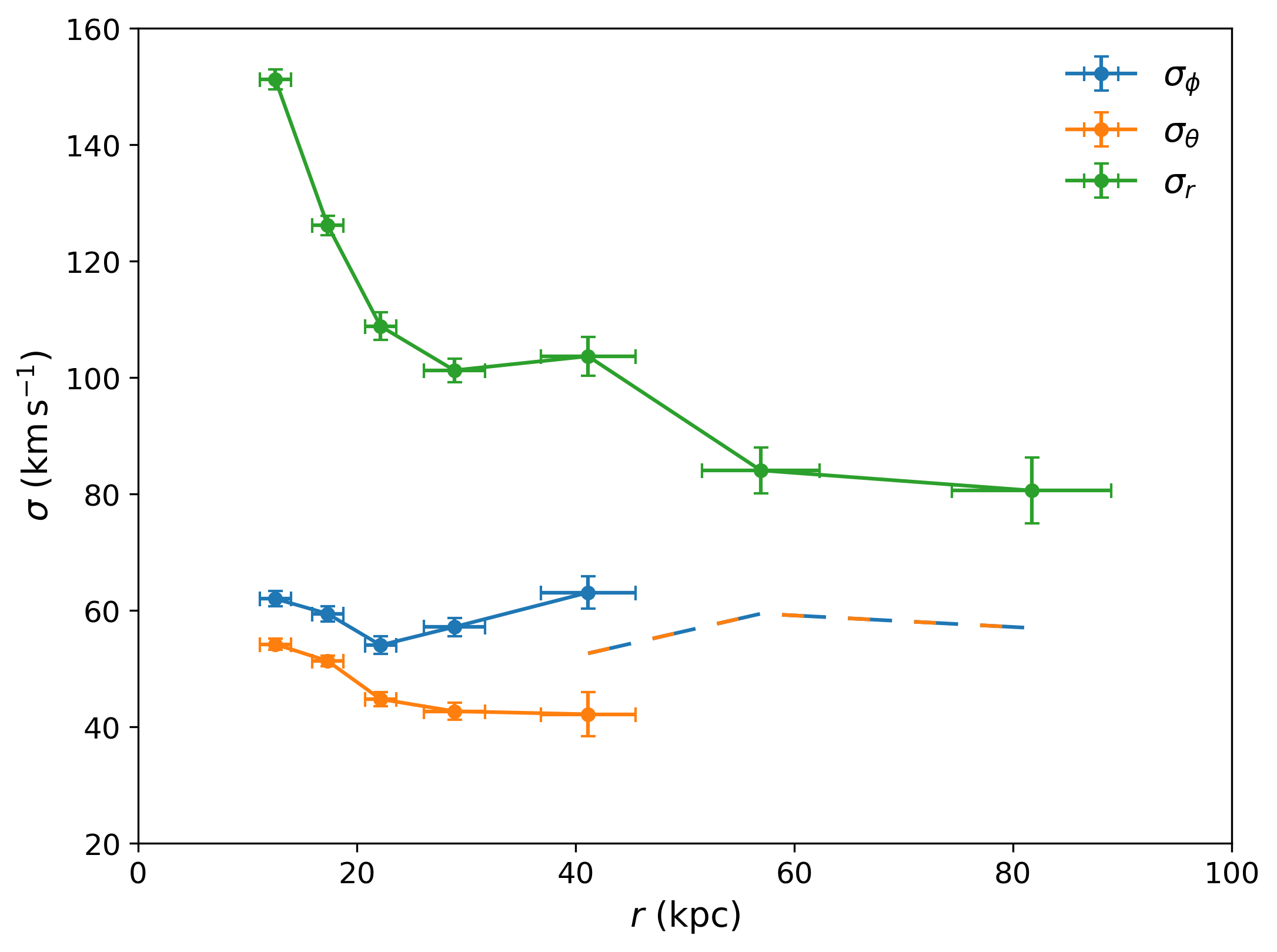}{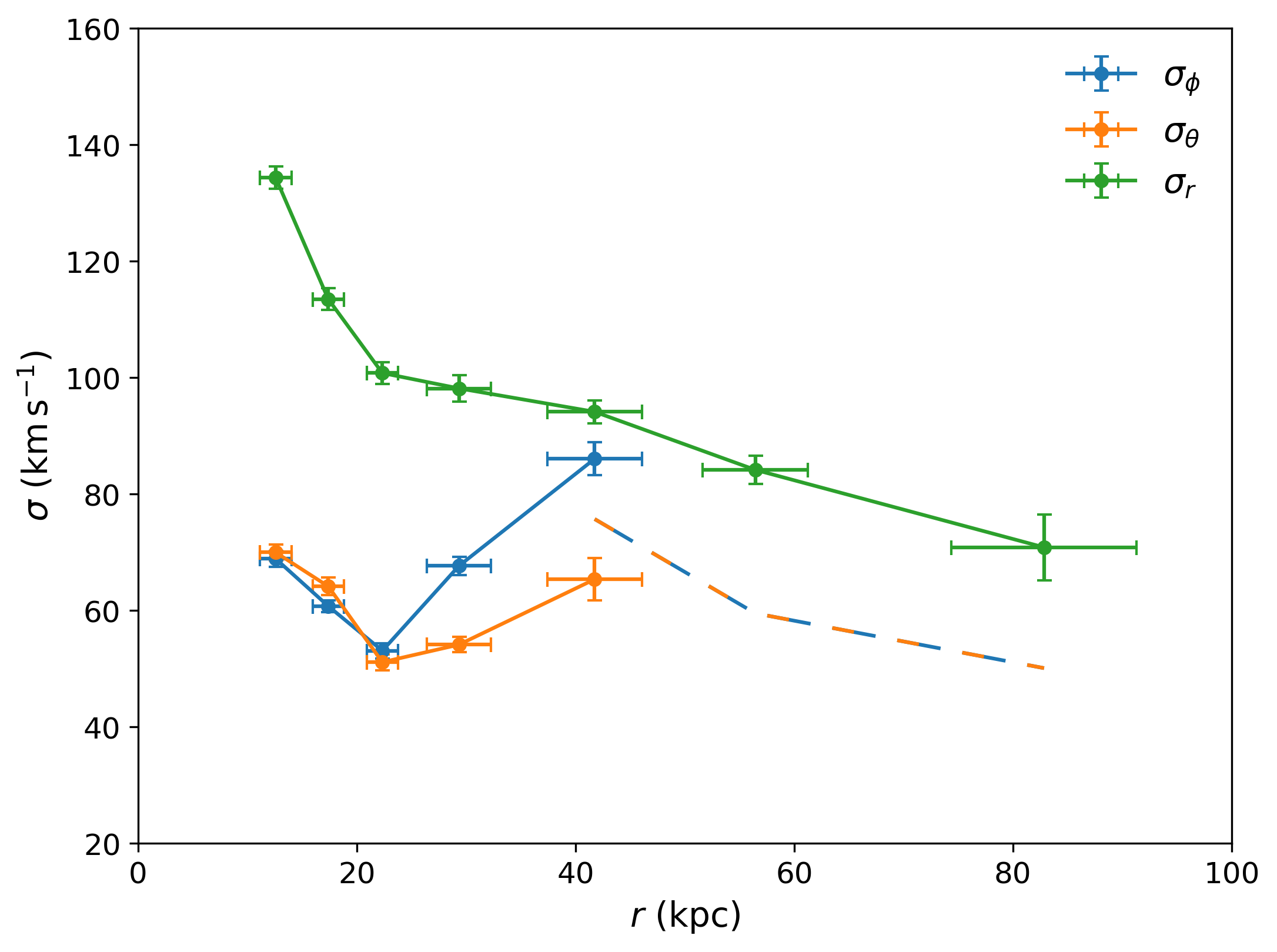}
\caption{Velocity dispersion profiles in Galactocentric spherical coordinates for the RRL (left) and BHB (right) samples. The green, blue, and yellow solid lines represent the radial ($\sigma_r$), polar ($\sigma_\theta$), and azimuthal ($\sigma_\phi$) velocity dispersions, respectively. The blue and yellow dashed lines show the corresponding tangential velocity dispersions inferred from the radial velocity dispersion and the velocity anisotropy parameter $\beta$.}
\label{fig:dispersion}
\end{figure*}

\subsection{Method 1: Spherical Jeans Equation \label{subsec:3.1}}

We use the spherical Jeans equation for a non-rotating, stationary system of stars, to measure the Galactic mass profile $M(<r)$, where $M$ is the mass enclosed within the Galactocentric radius $r$:

\begin{equation}
M(<r) = -\frac{1}{G}(r^2\frac{d\sigma_r^2}{dr}+r((2+\alpha)\sigma_r^2-\sigma_\theta^2-\sigma_\phi^2))
\label{eq:Jeans}
\end{equation}

Here, $G$ is the gravitational constant, and the velocity dispersions in the radial and two transverse directions are given by $(\sigma_r, \sigma_\theta, \sigma_\phi)$. The density profile of the tracer stars is modeled as a power-law profile $\rho \propto r^\alpha$, where $\alpha$ is the power-law index. This assumption remains one of the important sources of uncertainty in our final mass estimates. A detailed discussion of the density profile used in this work is given in Section \ref{sec:2.2}.

The velocity dispersions are derived using the {\tt\string XDGMM} function from the {\tt\string astroML} module in {\tt\string Python}. This function is based on the extreme-deconvolution method from \citet{Bovy2011}, using a maximum likelihood density estimation technique to model the error-deconvolved distribution as a Gaussian component. A key strength of this method is that the likelihood treats not only the mean velocities of each component but also the velocity uncertainties of every individual star. The velocity dispersions at different radii are shown in Figure~\ref{fig:dispersion}.

Due to the significant errors in \textit{Gaia} DR3 proper motion measurements at large distances (see Figure~\ref{fig:pm}), we decided not to utilize proper motion for stars beyond 50\,kpc. Instead, we adopt anisotropy parameter $\beta=0.5$ from \citet{RRL3} and \citet{Bird2021} for the RRLs and BHBs beyond 50\,kpc. The velocity dispersion in the transverse directions $\sigma_\phi$ and $\sigma_\theta$ were then derived from the anisotropy parameter $\beta$ and the velocity dispersion in the radial direction $\sigma_r$.

The spherical Jeans equation involves the radial gradient of the radial velocity dispersion squared, $\frac{\mathrm{d}\sigma_r^2}{\mathrm{d}r}$. To evaluate this term, we fit the radial velocity dispersion profile, $\sigma_r(r)$, with an empirical exponential function (see \citealt{Bird}): 
\begin{equation}
\sigma_r = \sigma_0\,\mathrm{exp}\left[-\frac{r}{h_r}\right].
\label{eq:Jeans1}
\end{equation}
The corresponding radial gradient, $\frac{\mathrm{d}\sigma_r^2}{\mathrm{d}r}$, is then obtained analytically by differentiating the fitted function:
\begin{equation}
\frac{\mathrm{d}\sigma_r^2}{\mathrm{d}r} = -\frac{2\sigma_0^2}{h_r}\,\mathrm{exp}\left[-\frac{2r}{h_r}\right].
\label{eq:Jeans2}
\end{equation}

\begin{figure*}
\centering
\plottwo{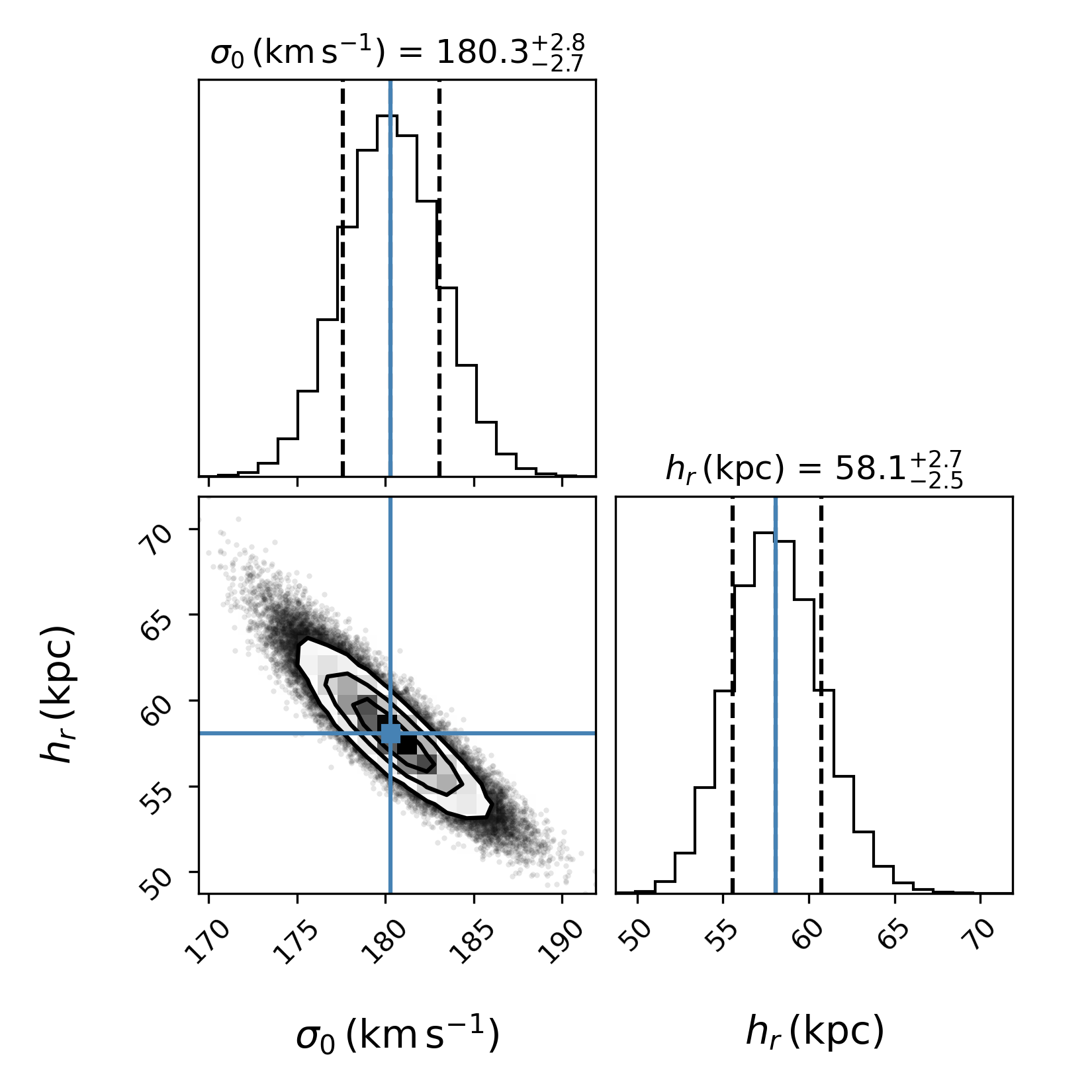}{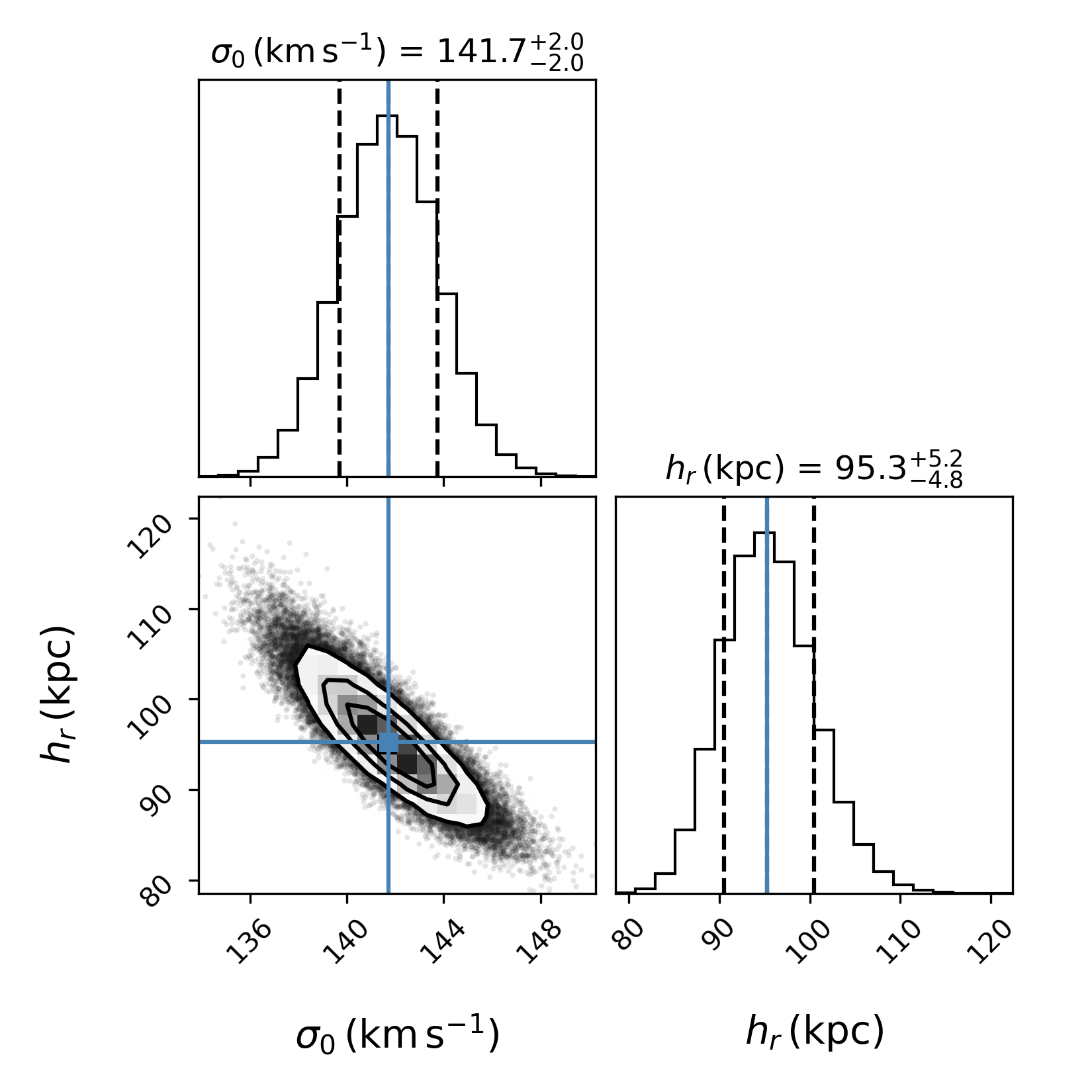}
\caption{Posterior distributions of $\sigma_0$ and $h_r$ obtained from Bayesian MCMC fitting of the radial velocity dispersion profiles for the RRL (left) and BHB (right) samples. The blue solid and black dashed lines indicate the median (50th percentile) and the 16th and 84th percentiles of the posterior distributions, respectively.}
\label{fig:mcmc}
\end{figure*}

We perform a Bayesian Markov chain Monte Carlo (MCMC) procedure using the {\tt\string Python} package {\tt\string emcee} to find the best fit for $\sigma_0$ and $h_r$. We use 1,000 walkers, 900 runs, and 100 runs for the burn-in phase, with minimal priors requiring $0<\sigma_0<1000$\,km\,s$^{-1}$ and $0<h_r<1000$\,kpc. The MCMC result for the RRLs and BHBs are shown in Figure~\ref{fig:mcmc}. We note that the parametric $\sigma_r$ is only used in the gradient term, the velocity dispersions in the Jeans equation are measured directly from extreme-deconvolution.

\subsection{Method 2: Tracer Mass Estimator \label{subsec:3.2}}

Estimation of the mass of a gravitationally bound system typically involves calculating a weighted average of the velocities and distances. Traditional mass estimators are based on the virial theorem with pioneering work by \citet{Zwicky1933}. Virial mass estimators take the form of two separate averages for distance and velocity. However, \citet{Bahcall1981} showed that virial mass estimators are inefficient, biased, and inconsistent, and replaced them with the projected mass estimator (PME).

In this work, we adopt the tracer mass estimator (TME) of \citet{Evans2011} to determine the enclosed mass of the Milky Way:

\begin{equation}
M(<r) = \frac{r^{0.5}(\gamma-\alpha-2\beta)}{GN}\sum^{N}_{i=1}r_i^{0.5}(V_{r,i}-\langle V_r \rangle)^2,
\label{eq:TME}
\end{equation}

where $M(<r)$ is the mass enclosed within Galactocentric radius $r$, $G$ is the gravitational constant, and $N$ is the number of tracer stars. For the $i$th tracer, $r_i$ and $V_{r,i}$ denote its Galactocentric distance and radial velocity, respectively, while $\langle V_r \rangle$ represents the mean Galactocentric radial velocity of the tracer sample. The parameters $\gamma$ and $\alpha$ characterize the power-law slopes of the Galactic gravitational potential and the tracer density profile, respectively, with the latter following the same prescription as adopted in the Jeans analysis. The parameter $\beta$ describes the velocity anisotropy of the tracer population and is defined as

\begin{equation}
\beta = 1-\frac{\sigma_\theta^2+\sigma_\phi^2}{2\sigma_r^2},
\label{eq:beta}
\end{equation}

where $\sigma_r$, $\sigma_\theta$, and $\sigma_\phi$ are the velocity dispersions in the radial, polar, and azimuthal directions, respectively, in Galactocentric spherical coordinates.

The TME generalizes the projected mass estimator by allowing the spatial distribution of tracer populations to differ from the underlying mass distribution. It has been extensively investigated and validated using simulated galaxies from $N$-body simulations \citep{Evans2003,Evans2011}.

By examining NFW potentials spanning a wide range of virial radii and concentrations, \citet{Watkins2010} demonstrated that the gravitational potentials of massive galaxies such as the Milky Way can be reasonably approximated by a scale-free power-law potential with an index of $\gamma=0.5$. We therefore adopt $\gamma=0.5$, consistent with the NFW-like dark matter halo generally assumed for the Milky Way.

Like the spherical Jeans equation, the TME assumes a spherically symmetric, non-rotating tracer population in dynamical equilibrium. These assumptions are expected to hold approximately for our halo samples, although departures from equilibrium may arise in the outer halo due to perturbations induced by the Large Magellanic Cloud (LMC). To mitigate the effects of such bulk motions, we subtract the mean Galactocentric radial velocity, $\langle V_r \rangle$, from the individual radial velocities when applying the TME, following \citet{Erkal2020} and \citet{Deason2021}. This correction reduces the contribution of coherent radial motions to the inferred mass, although residual systematic effects associated with the LMC may remain.

We estimate the statistical uncertainties in the enclosed mass profile using bootstrap resampling. For each radial bin, we generate 100 bootstrap realizations of the tracer sample and recompute the enclosed mass for each realization. We adopt half the difference between the 84th and 16th percentiles of the resulting mass distribution as the $1\sigma$ uncertainty.

\subsection{Virial Mass Estimation \label{sec:3.3}}

The enclosed mass profiles derived from the Jeans equation and the TME represent the total gravitational mass of the Milky Way, including both baryonic and dark matter components. To constrain the dark matter halo, we model the Galactic mass distribution as the sum of three components: a stellar bulge, a stellar disk, and a dark matter halo. The bulge and disk parameters are fixed, while the dark matter halo parameters are determined by fitting the observed enclosed mass profiles.

\textbf{Bulge.}
We model the Galactic bulge using a Hernquist density profile \citep{Hernquist},

\begin{equation}
\rho_{\rm bulge}(r) =
\frac{M_{\rm bulge}}{2\pi}
\frac{a_{\rm b}}{r(r+a_{\rm b})^3},
\label{eq:bulge}
\end{equation}

where $M_{\rm bulge}$ and $a_{\rm b}$ are the total bulge mass and scale radius, respectively. We adopt $M_{\rm bulge}=5\times10^{9}\,\mathrm{M}_\odot$ and $a_{\rm b}=1.9$\,kpc.

\textbf{Disk.}
The Galactic disk is represented by a Miyamoto--Nagai potential \citep{Miyamoto},

\begin{equation}
\Phi_{\rm disk}(R,z) =
-\frac{GM_{\rm disk}}
{\sqrt{R^2+\left(a_{\rm d}+\sqrt{z^2+b_{\rm d}^2}\right)^2}},
\label{eq:disk}
\end{equation}

where $R$ and $z$ are the cylindrical Galactocentric coordinates, and $M_{\rm disk}$, $a_{\rm d}$, and $b_{\rm d}$ denote the disk mass, scale length, and scale height, respectively. We adopt $M_{\rm disk}=6.8\times10^{10}\,\mathrm{M}_\odot$, $a_{\rm d}=3$\,kpc, and $b_{\rm d}=0.28$\,kpc.

\textbf{Dark matter halo.}
We describe the Galactic dark matter halo using a spherical Navarro--Frenk--White (NFW) density profile \citep{NFW1,NFW2},

\begin{equation}
\rho_{\rm halo}(r) =
\frac{\rho_{\rm s}}
{(r/r_{\rm s})(1+r/r_{\rm s})^2},
\label{eq:NFW}
\end{equation}

where $\rho_{\rm s}$ and $r_{\rm s}$ are the characteristic density and scale radius of the halo, respectively. The scale radius is related to the virial radius $r_{200}$ and concentration parameter $c$ through $r_{\rm s}=r_{200}/c$. The virial radius is defined as the radius within which the mean enclosed density is 200 times the critical density of the Universe, $\rho_{\rm crit}=3H^2/(8\pi G)$, where $H$ is the Hubble constant and $G$ is the gravitational constant. The corresponding virial mass is

\begin{equation}
M_{200} =
\frac{4\pi}{3}(200\rho_{\rm crit})r_{200}^3.
\label{eq:M200}
\end{equation}

The characteristic density $\rho_{\rm s}$ can therefore be expressed in terms of $M_{200}$ and $c$ as

\begin{equation}
\rho_{\rm s} =
\frac{M_{200}}
{4\pi r_{\rm s}^3
\left[\ln(1+c)-c/(1+c)\right]}.
\label{eq:rhos}
\end{equation}

We implement these Galactic components using the \texttt{galpy} package \citep{galpy}. We adopt a Hubble constant of $H=67.8$\,km\,s$^{-1}$\,Mpc$^{-1}$ \citep{Hubble} to define the critical density of the Universe and the corresponding virial mass $M_{200}$.

To constrain the dark matter halo, we subtract the modeled baryonic contribution from the observed enclosed mass profiles and fit the residual profiles with the NFW model. We treat the virial radius $r_{200}$ and concentration $c$ as free parameters and explore their posterior distributions using the MCMC sampler implemented in the \texttt{emcee} package. We employ 500 walkers, each evolved for 600 steps, with the first 200 steps discarded as burn-in. Uniform priors of $100<r_{200}/{\rm kpc}<400$ and $0<c<30$ are adopted. The resulting posterior distributions are used to constrain the virial mass $M_{200}$ and halo concentration $c$.

\section{Results and Validation \label{sec:4}}

\subsection{Results \label{sec:4.1}}

\begin{figure*}
\centering
\plottwo{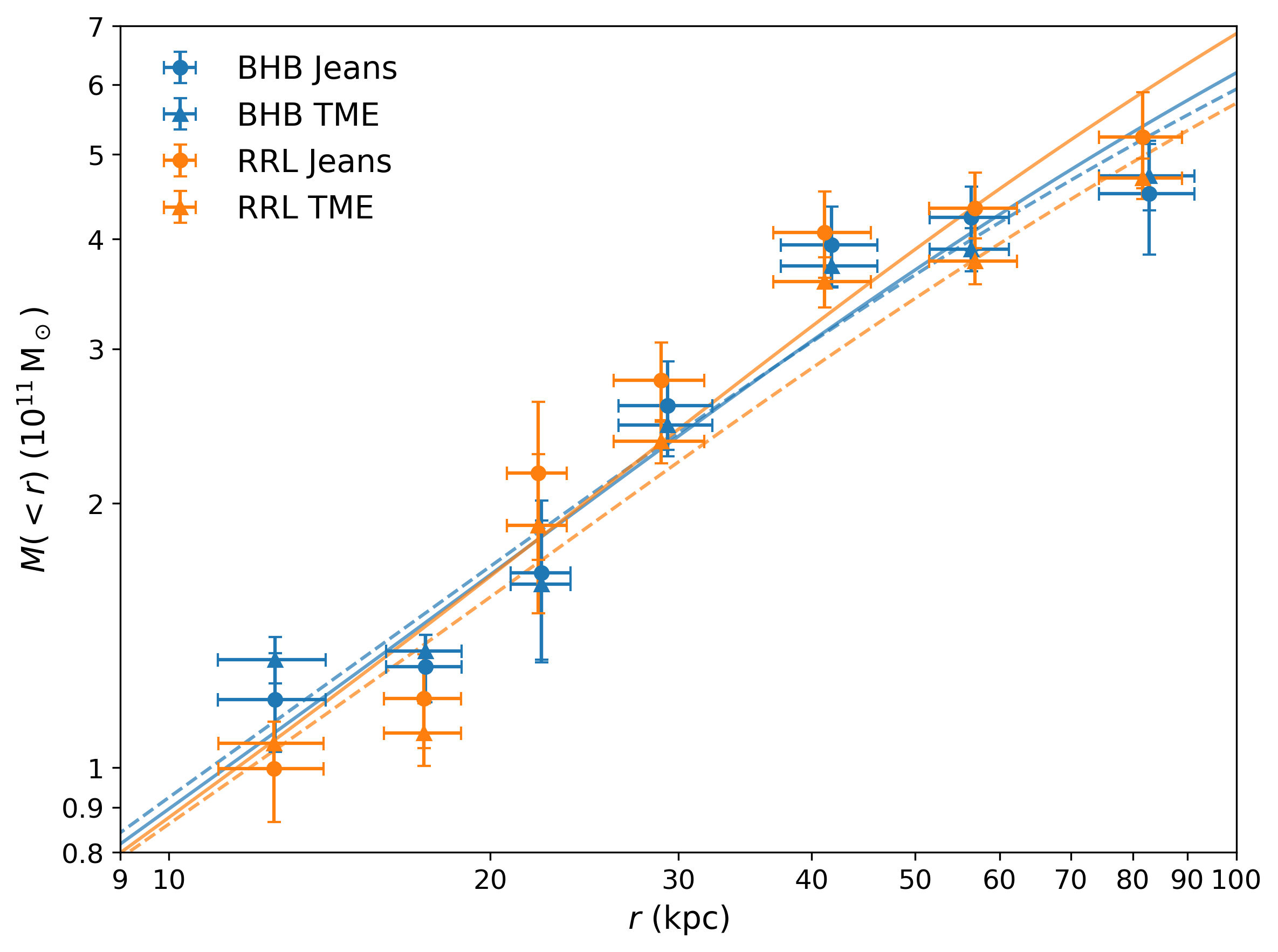}{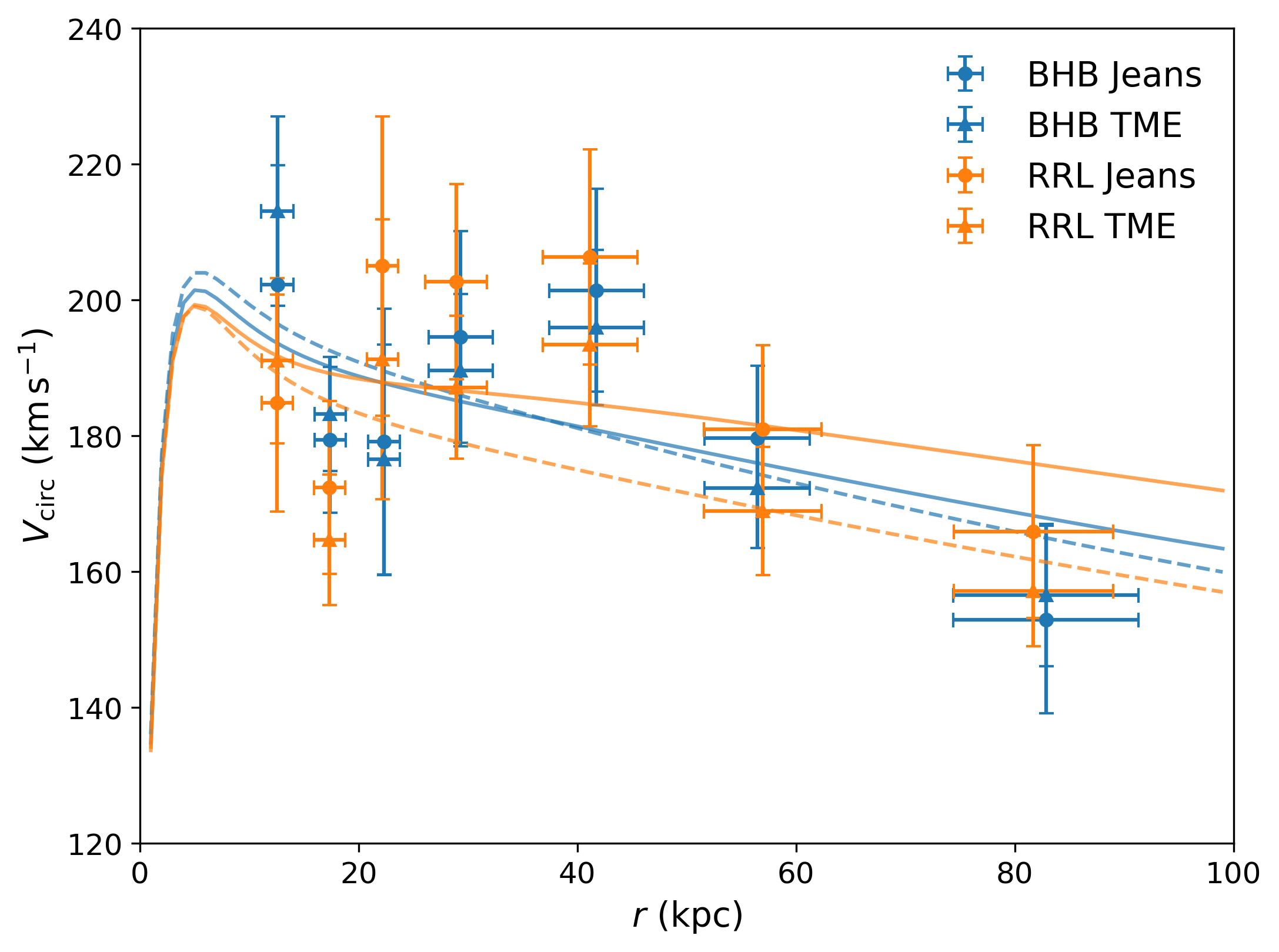}
\caption{Enclosed mass and circular velocity profiles of the Milky Way. The blue and orange data points show the enclosed mass measurements derived from the BHB and RRL samples, respectively. The solid and dashed curves represent the best-fitting NFW models based on the enclosed mass profiles inferred using the Jeans and TME methods, respectively.}
\label{fig:mass}
\end{figure*}

The enclosed mass measurements are summarized in Table \ref{tab:1}, while the corresponding enclosed mass and circular velocity profiles are presented in Figure~\ref{fig:mass}. The circular velocity is calculated as

\begin{equation}
v_{\rm circ}(r)=\sqrt{\frac{GM(<r)}{r}},
\label{eq:vcirc}
\end{equation}

where $M(<r)$ is the mass enclosed within Galactocentric radius $r$. The uncertainties shown in Figure~\ref{fig:mass} reflect only the statistical uncertainties associated with the measurements, which are typically $\sim10$--$30\%$, and do not account for systematic uncertainties arising from model assumptions.

\begin{figure*}
\centering
\plottwo{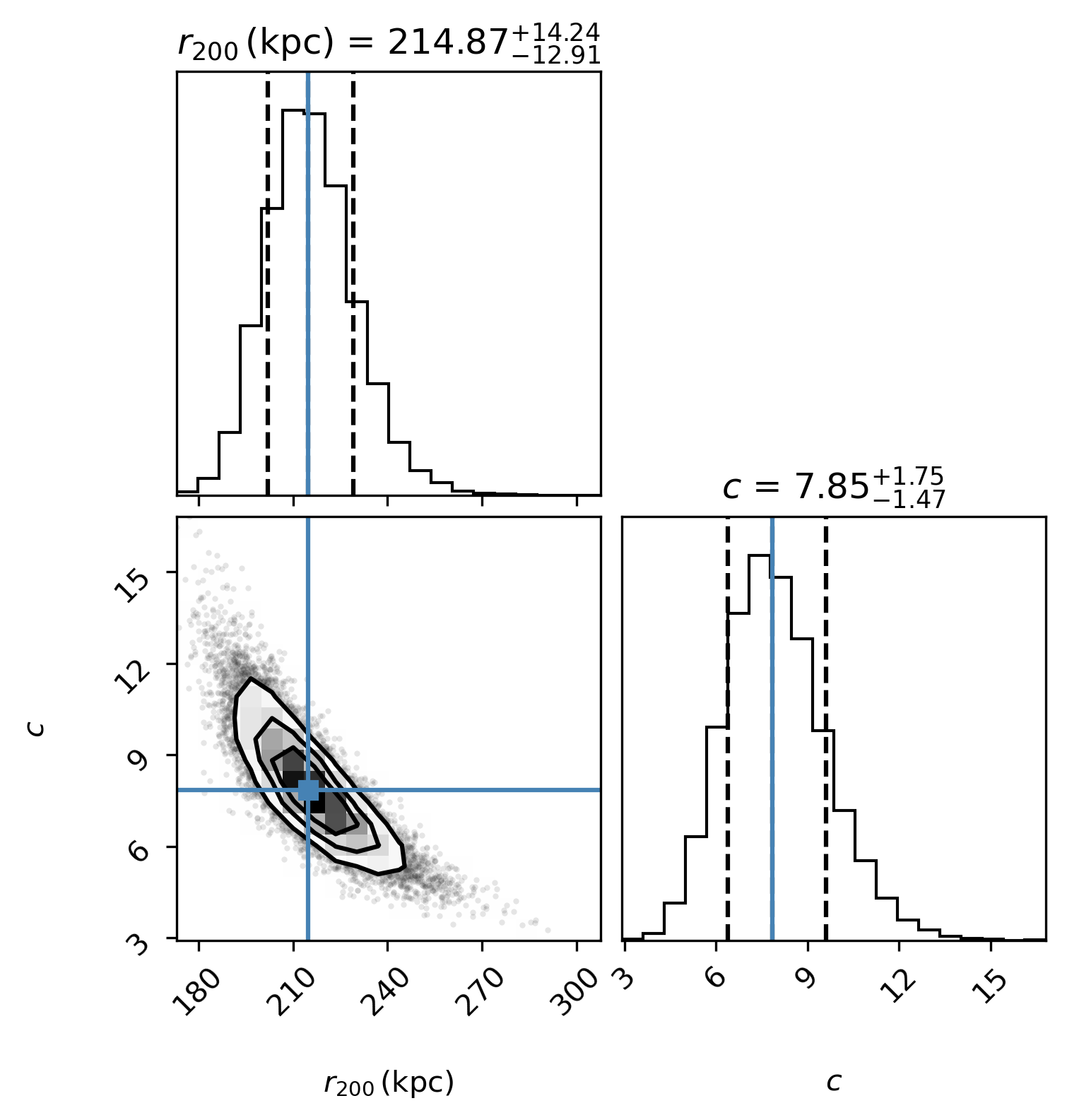}{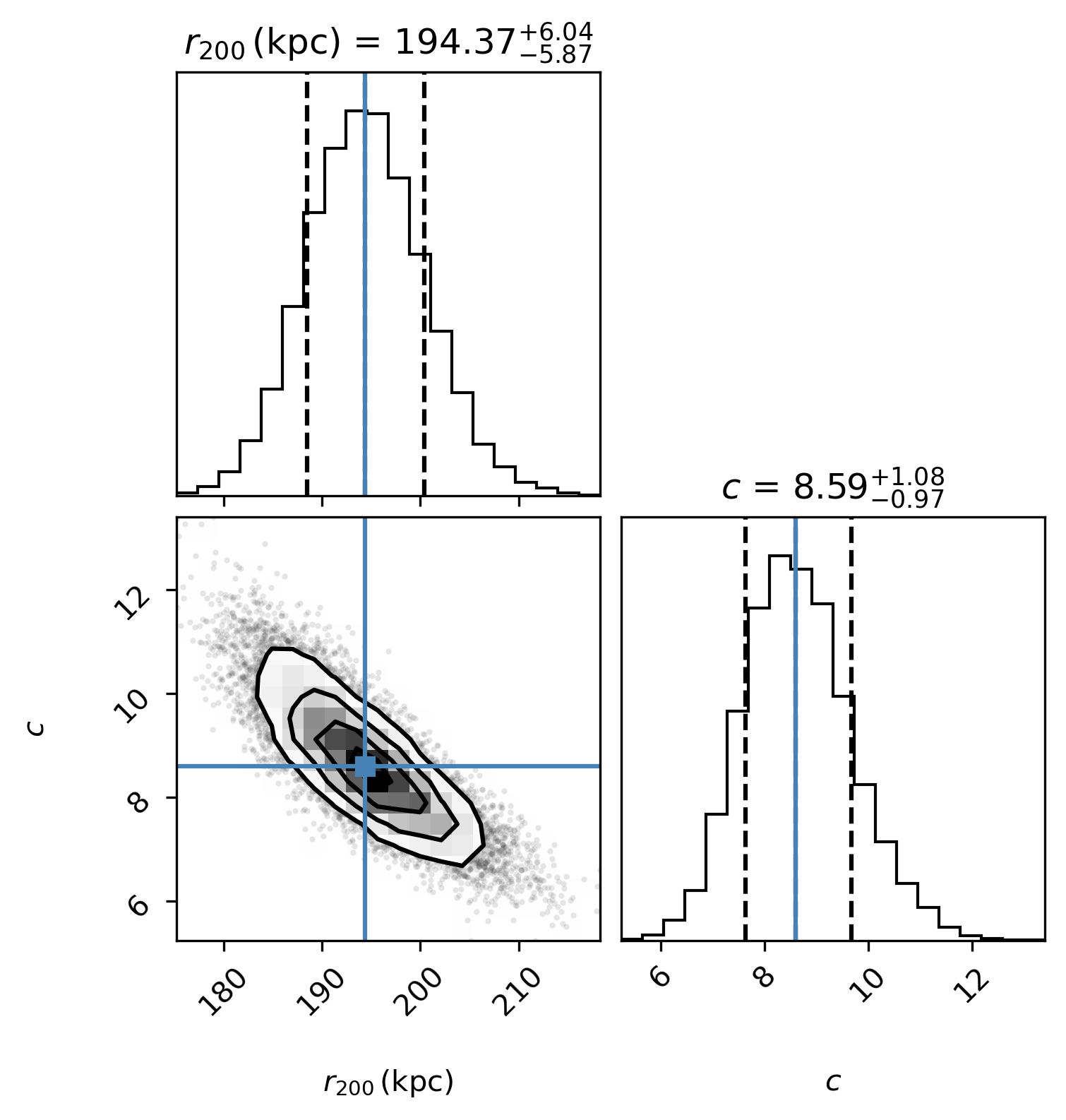}
\plottwo{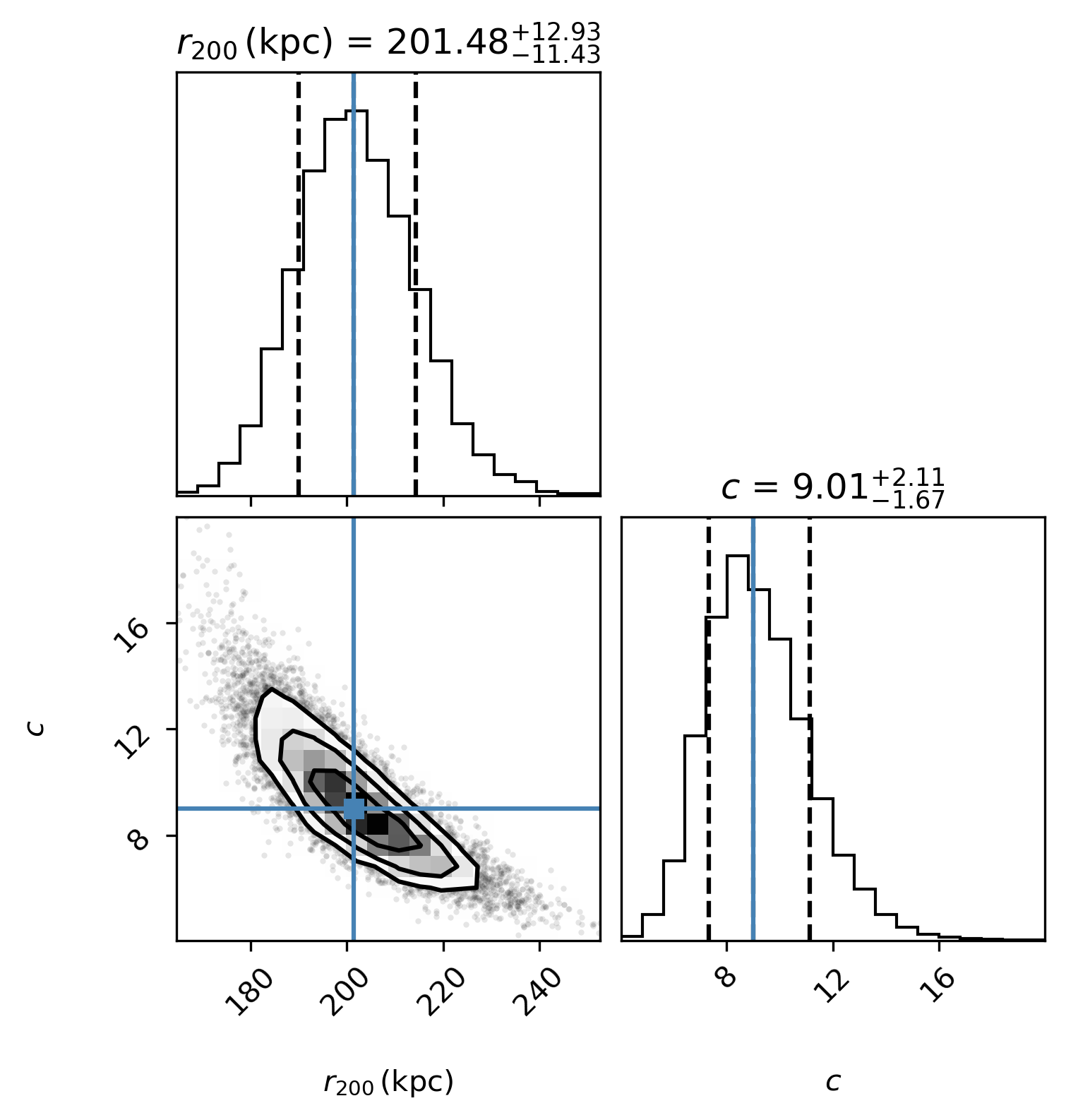}{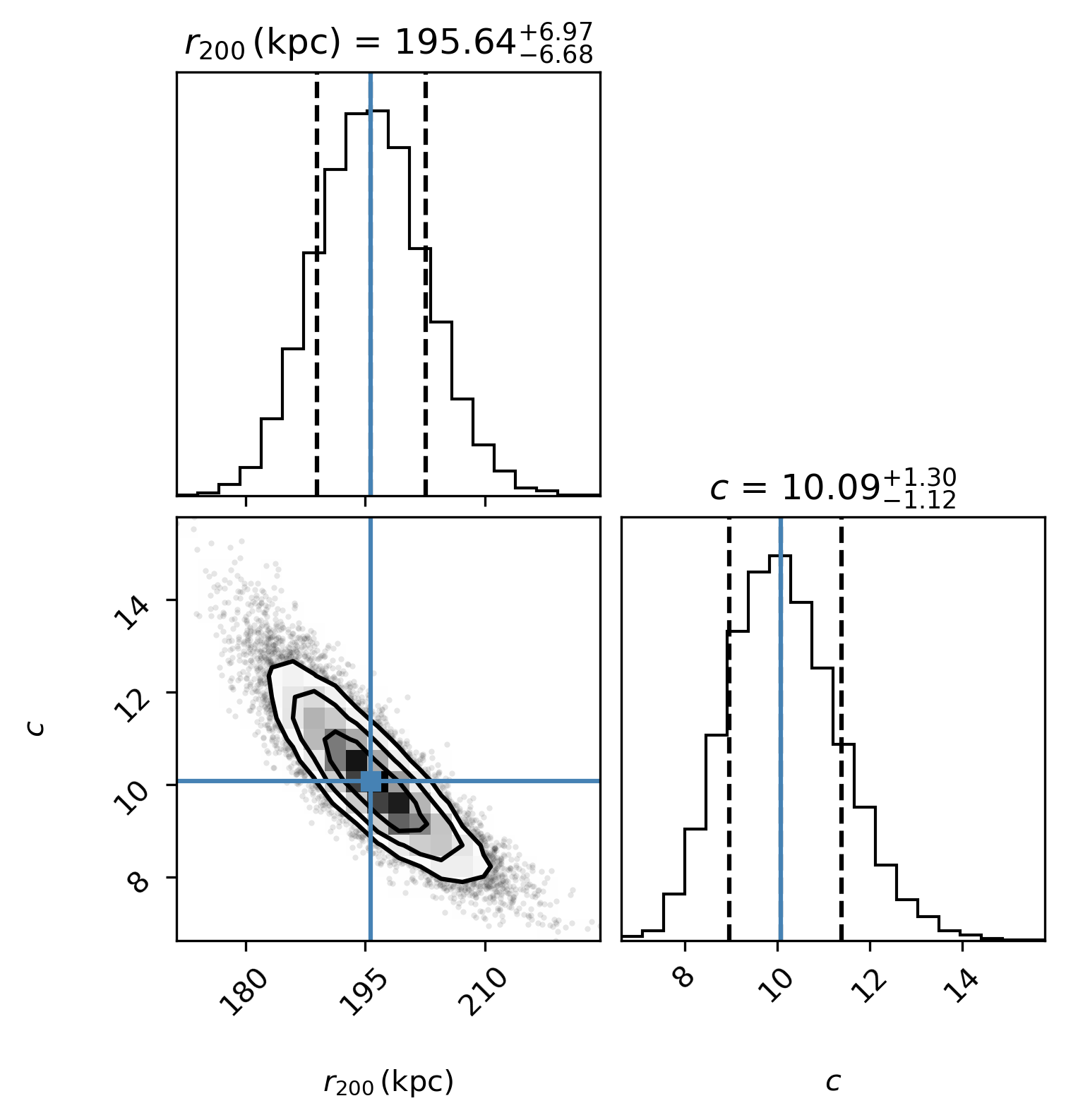}
\caption{Posterior distributions of the virial radius ($r_{200}$) and concentration ($c$) obtained from Bayesian MCMC fitting of the NFW halo model for the RRL (top) and BHB (bottom) samples, using the Jeans (left) and TME (right) methods. The blue solid and black dashed lines indicate the median (50th percentile) and the 16th and 84th percentiles of the posterior distributions, respectively.}
\label{fig:galpy}
\end{figure*}

The virial mass ($M_{200}$), virial radius ($r_{200}$), and concentration ($c$) inferred from the NFW halo fitting are summarized in Table \ref{tab:2}, with their posterior distributions shown in Figure~\ref{fig:galpy}. The resulting virial masses are $M_{200}=8.12^{+0.72}_{-0.65}\times10^{11}\,\mathrm{M}_\odot$ and $M_{200}=8.13^{+0.78}_{-0.67}\times10^{11}\,\mathrm{M}_\odot$ for the RRL and BHB samples, respectively. These two independent tracer populations yield remarkably consistent estimates of the Milky Way's virial mass.

\begin{table}
\begin{flushleft}
\caption{Enclosed mass profiles for the Milky Way.}
\begin{tabular}{lcc}
\hline
\textbf{Method(star type)} & \textbf{$<r^\dagger(\mathrm{kpc})$} & \textbf{$M(<r)(\times10^{11}\mathrm{M}_\odot)$} \\
\hline
Jeans (RRL) &81.7&$5.23\pm0.66$\\
TME (RRL) &81.7&$4.69\pm0.25$\\
Jeans (BHB) &82.8&$4.51\pm0.67$\\
TME (BHB) &82.8&$4.72\pm0.41$\\
\hline
\end{tabular}
\vskip 0.5cm
$\dagger$ Median Galactocentric radius of the stars in the outermost radial bin.
\label{tab:1}
\end{flushleft}
\end{table}

\begin{table}
\begin{flushleft}
\setlength{\tabcolsep}{3pt} 
\caption{Best-fitting parameters for the Milky Way.}
\begin{tabular}{lccc}
\hline
\textbf{Method(star type)} & \textbf{$M_{200}(\times10^{11}\mathrm{M}_\odot)$} & \textbf{$r_{200}(\mathrm{kpc})$} & \textbf{$c$} \\
\hline
Jeans (RRL)&$10.59^{+2.25^\dagger}_{-1.81}$&$215.1^{+14.3}_{-13.0}$&$7.8^{+1.8}_{-1.5}$\\
TME (RRL)&$7.80^{+0.76}_{-0.69}$&$194.3^{+6.2}_{-5.9}$&$8.6^{+1.1}_{-1.0}$\\
Jeans (BHB)&$8.71^{+1.73}_{-1.38}$&$201.6^{+12.5}_{-11.2}$&$9.0^{+2.0}_{-1.7}$\\
TME (BHB)&$7.97^{+0.87}_{-0.77}$&$195.6^{+6.9}_{-6.6}$&$10.1^{+1.3}_{-1.1}$\\
\hline
\end{tabular}
\vskip 0.5cm
$\dagger$ All parameter estimates are obtained from Bayesian MCMC analyses. We adopt the median (50th percentile) of the posterior distribution as the best-fit value, with the 16th and 84th percentiles defining the corresponding $1\sigma$ credible interval.
\label{tab:2}
\end{flushleft}
\end{table}

\begin{table}
\begin{flushleft}
\setlength{\tabcolsep}{10pt}
\caption{Best-fitting virial mass for the Auriga galaxies.}
\begin{tabular}{lccc}
\hline
\textbf{Method(star type)} & \textbf{$M_{200}(\mathrm{Au6})$} & \textbf{$M_{200}(\mathrm{Au16})$} \\
&\textbf{$(\times10^{12}\mathrm{M}_\odot)$}&\textbf{$(\times10^{12}\mathrm{M}_\odot)$}\\
\hline
True Mass&$1.01$&$1.50$\\
Jeans (RRL)&$0.93^{+0.09^\dagger}_{-0.09}$&$1.46^{+0.14}_{-0.12}$\\
TME (RRL)&$0.75^{+0.04}_{-0.04}$&$1.43^{+0.08}_{-0.08}$\\
Jeans (BHB)&$0.87^{+0.10}_{-0.08}$&$1.39^{+0.18}_{-0.15}$\\
TME (BHB)&$0.85^{+0.05}_{-0.05}$&$1.41^{+0.10}_{-0.09}$\\
\hline
\end{tabular}
\vskip 0.5cm
$\dagger$ All estimates are derived from Bayesian MCMC fitting. We adopt the 50th percentile as the best-fit value, and the 16th and 84th percentiles as the $1\sigma$ random uncertainties.
\label{tab:3}
\end{flushleft}
\end{table}

\begin{table}
\begin{flushleft}
\setlength{\tabcolsep}{3pt} 
\caption{Corrected virial mass for the Milky Way.}
\begin{tabular}{lcc}
\hline
\textbf{Method(star type)} & \textbf{$M_{200}(\times10^{11}\mathrm{M}_\odot)$} & \textbf{Correction} \\
\hline
Jeans (RRL)&$11.17^{+2.48^\dagger}_{-2.04}$ & $+5.2\%\pm6.2\%$\\
TME (RRL)&$9.54^{+1.00}_{-0.92}$ & $+18.2\%\pm3.2\%$\\
Jeans (BHB)&$9.82^{+2.09}_{-1.73}$ & $+11.3\%\pm6.9\%$\\
TME (BHB)&$9.07^{+1.07}_{-0.96}$ & $+12.1\%\pm3.9\%$\\
\hline
\end{tabular}
\vskip 0.5cm
$\dagger$ The uncertainty in the corrected virial masses includes both the random errors arising from measurement and the systematic errors.
\label{tab:4}
\end{flushleft}
\end{table}

\begin{figure*}
\centering
\plotone{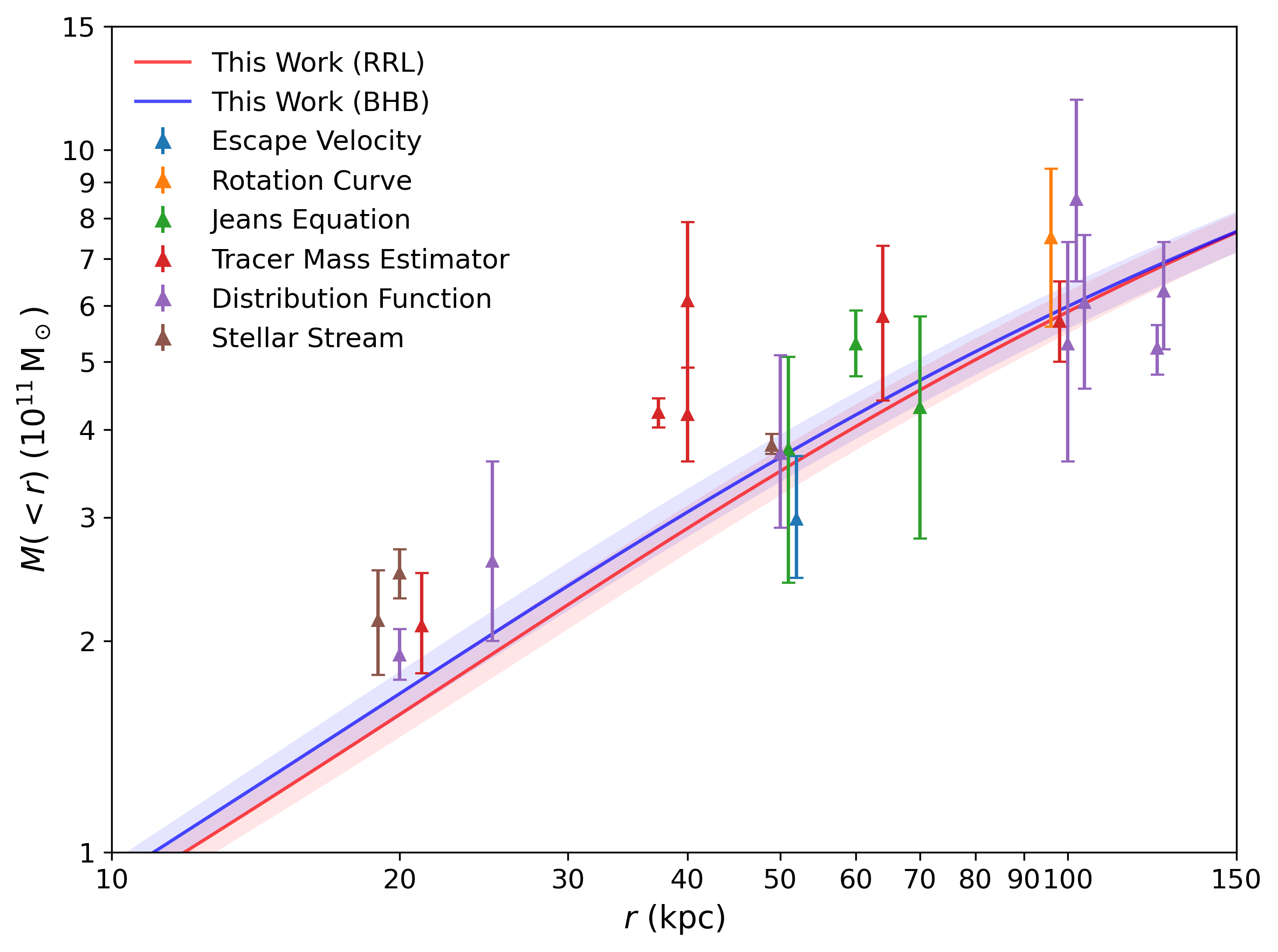}

\caption{Comparison of the enclosed mass profiles derived in this work with previous measurements from the literature. The blue and red curves show the mass profiles obtained from the BHB and RRL samples, respectively, with the shaded regions indicating the corresponding 16th--84th percentile credible intervals.
The colored data points represent literature measurements compiled from
\citet{Wang2020, Williams2017, Bajkova2016, Ablimit2017, Bird, Zhang2025, Watkins2019, Fritz2020, Sun2023, DESIMass, Eadie2016, Eadie2017, Posti2019, Eadie2019, Vasiliev2019, Callingham2019, Deason2021, Kupper2015, Malhan2019, Erkal2019}.
The colors of the data points distinguish the different mass estimation methods employed in these studies.}
\label{fig:histo2}
\end{figure*}

\subsection{Method Validation \label{sec:4.2}}

Both the spherical Jeans equation and the TME rely on the assumptions of dynamical equilibrium and spherical symmetry. In addition, our determination of the virial mass assumes that the Galactic dark matter halo follows an NFW profile. Departures from these assumptions may introduce systematic biases into the inferred mass of the Milky Way.

To assess these potential biases, we apply our mass estimation methods to mock stellar halo samples drawn from the Auriga cosmological simulations \citep{Auriga}. We select two Milky Way-like galaxies, Au-6 and Au-16, which exhibit different structural properties. Au-6 hosts a disk resembling that of the Milky Way, whereas Au-16 has a more extended disk and more massive substructures. For both simulated galaxies, the mock Sun is placed at a Galactocentric radius of 8\,kpc, 20\,pc above the Galactic mid-plane, and at an azimuthal angle of $30^\circ$ relative to the Galactic bar.

We construct mock RRL and BHB samples from the Auriga stellar catalogues by applying the following selection criteria:

\begin{gather}
\mathrm{Age} > 10\,\mathrm{Gyr}, \notag \\
2.5 < \log g < 3.0, \notag \\
0.4 < M_G < 0.85, \notag \\
5500 < T_{\rm eff}/{\rm K} < 7300
\quad (\mathrm{RRL}), \notag \\
7300 < T_{\rm eff}/{\rm K} < 9750
\quad (\mathrm{BHB}),
\end{gather}

where $M_G$ is the absolute magnitude in the \textit{Gaia} $G$ band, $\log g$ is the logarithmic surface gravity, and $T_{\rm eff}$ is the effective temperature. We use the true distances and velocities of the simulated stars and introduce Gaussian measurement uncertainties calibrated to reproduce the typical uncertainties of the observational samples.

We then apply the same selection procedures as those used for the observational data (Section~\ref{sec:2.1}) to remove disk contaminants and identified halo substructures. From the remaining stars, we randomly select 10,000 tracers to construct mock samples with sizes comparable to those of our observational samples.

We analyze the mock samples using the same procedures adopted for the observations. Specifically, we deriv the enclosed mass profiles using both the Jeans equation and the TME, and subsequently fit NFW halo models using Bayesian MCMC to infer the virial masses. In Table~\ref{tab:3}, we compare the recovered masses with the true values from the simulations to quantify the systematic biases introduced by our methodology. As summarized in Table~\ref{tab:4}, the inferred virial masses differ from the true values by $15.5\%\pm2.8\%$ for the RRLs, and $11.9\%\pm3.4\%$ for the BHBs, respectively. After correcting for these systematic errors, the Milky Way virial masses are $M_{200} = 9.79^{+0.93}_{-0.84}\times10^{11}\,\mathrm{M}_\odot$ from RRLs and $M_{200} = 9.24^{+0.95}_{-0.84}\times10^{11}\,\mathrm{M}_\odot$ from BHBs.

\section{Discussion and Conclusion \label{sec:5}}

\subsection{Sources of Uncertainty \label{sec:5.1}}

Previous measurements of the Milky Way mass have been limited by substantial statistical and systematic uncertainties, partly due to the limited number of available halo tracers. Recent large-scale spectroscopic surveys, together with the precise astrometric measurements from \textit{Gaia}, have greatly increased the number of halo stars with reliable kinematic information, substantially reducing statistical uncertainties. As a result, systematic uncertainties arising from the assumptions underlying dynamical mass estimators have become increasingly important. In particular, the assumptions of spherical symmetry and dynamical equilibrium may not be fully satisfied by the Milky Way halo, even after known stellar streams and substructures have been removed.

In this section, we discuss the potential systematic uncertainties associated with the \textit{Gaia}--Enceladus--Sausage (GES), the Large Magellanic Cloud (LMC), and departures from spherical symmetry and dynamical equilibrium. These effects are distinct from the statistical uncertainties discussed in Section~\ref{sec:3}.

The \textit{Gaia}--Enceladus--Sausage \citep[GES;][]{GES1,GES2} is a major accreted component of the inner Galactic halo and has a substantial impact on the velocity distribution of halo stars. Although we have excluded identified stellar streams and substructures, GES debris is widely distributed throughout the inner halo and cannot be reliably separated from the remaining halo population using kinematic information alone. Its strongly radially anisotropic velocity distribution may therefore affect our dynamical mass estimates. A more comprehensive treatment would require explicitly incorporating the GES component into the dynamical model, which is beyond the scope of this work.

The gravitational influence of the LMC provides another important source of systematic uncertainty, particularly in the outer Galactic halo. The infall of the LMC induces a reflex motion of the Milky Way and a spatially dependent kinematic response of halo stars, with increasingly significant effects at large Galactocentric distances, particularly beyond $\sim50$\,kpc \citep{Erkal2020,Petersen2020}. Previous studies have shown that neglecting these perturbations can lead to an overestimation of the Milky Way mass by up to $\sim25\%$, depending on the tracer population and spatial coverage \citep{Erkal2020}. To mitigate this effect, we subtract the mean velocity of the tracer population when calculating the velocity dispersions in the Jeans analysis and the radial velocity term in the TME, following \citet{Erkal2020}. We find that this correction changes our inferred Milky Way mass by approximately $6$--$7\%$. The relatively modest impact compared with that reported by \citet{Erkal2020} may be partly attributed to the spatial distribution of our halo tracers, which predominantly occupy regions less strongly affected by the LMC \citep{Deason2021}. Nevertheless, subtracting the mean velocity does not fully account for the spatially varying perturbations induced by the LMC, and residual systematic uncertainties may remain.

Both the Jeans and TME methods rely on assumptions of spherical symmetry and dynamical equilibrium. Departures from these assumptions, including halo asphericity, coherent bulk motions, and incomplete phase mixing, can introduce biases into the inferred mass profiles. In addition, our implementation of the TME assumes a scale-free gravitational potential with a power-law index of $\gamma=0.5$ \citep{Evans2011,Watkins2010}, while the subsequent determination of the virial mass with both methods assumes an NFW dark matter halo profile. These approximations introduce additional model-dependent uncertainties. Based on our tests with the Auriga simulations presented in Section~\ref{sec:4.2}, we estimate that the combined systematic effects associated with departures from the adopted dynamical and structural assumptions are $15.5\%\pm2.8\%$ for the RRLs, and $11.9\%\pm3.4\%$ for the BHBs, respectively.

The Jeans and TME methods also differ in their treatment of the tracer kinematics. The spherical Jeans equation explicitly incorporates the radial velocity dispersion profile and the velocity anisotropy parameter, whereas the TME uses the radial velocities of individual tracers, with the tangential velocity information entering through the anisotropy parameter $\beta$. Consequently, the two methods are subject to different sources of uncertainty. In particular, the Jeans method requires the radial gradient of the radial velocity dispersion squared, which must be estimated from the observed dispersion profile. In this work, we model the radial velocity dispersion using an empirical exponential function, which provides a smooth representation of the observed profile and allows its gradient to be evaluated analytically. The TME avoids this additional parametrization but depends explicitly on the assumed form of the gravitational potential. The agreement between the mass estimates obtained from these two methods therefore provides a useful consistency check, although systematic effects common to both methods may persist.

\subsection{Comparison with literature \label{sec:5.2}}

To place our results in a broader context, we compare our mass estimates with previous measurements based on a variety of methods, including escape velocity, rotation curves, distribution function modeling, and dynamical mass estimators, using different tracers such as halo stars, stellar streams, and satellite galaxies. Figure~\ref{fig:histo2} compares our enclosed mass profiles with those reported in the literature at different Galactocentric radii. Our results are broadly consistent with most previous measurements within 150\,kpc.

Figure~\ref{fig:histo} presents a comparison of our virial mass estimates with those from previous studies. Our results are in good agreement with estimates based on escape velocities and the Jeans equation, and are also consistent, within the reported uncertainties, with some measurements derived from rotation curves, the tracer mass estimator, and distribution function modeling. Overall, our measurements favor a relatively low Milky Way virial mass and are broadly compatible with the majority of recent estimates.

\subsection{Conclusions \label{sec:5.3}}

In this work, we use the full six-dimensional (6D) phase-space information of RRL stars and BHB stars from LAMOST, SDSS, and DESI to constrain the mass of the Milky Way. After excluding stellar streams, other halo substructures, and Galactic disk contaminants, our final sample consists of approximately 7,550 RRLs and 8,230 BHBs, extending to Galactocentric distances of 100\,kpc. We estimate the Milky Way mass using both the spherical Jeans equation and the tracer mass estimator (TME) of \citet{Evans2011}, leveraging the large samples of these two halo tracers and their well-calibrated distance estimates.

For the RRL and BHB samples, we derive an enclosed mass profile of $M(<81.7\,\mathrm{kpc}) = (4.76 \pm 0.23)\times10^{11}\,\mathrm{M}_\odot$ (RRLs) and $M(<82.8\,\mathrm{kpc}) = (4.66 \pm 0.35)\times10^{11}\,\mathrm{M}_\odot$ (BHBs). Extrapolating our mass profiles beyond the extent of our data, we report the virial mass of the Milky Way to be $M_{200} = 8.12^{+0.72}_{-0.65}\times10^{11}\,\mathrm{M}_\odot$ (RRLs) and $M_{200} = 8.13^{+0.78}_{-0.67}\times10^{11}\,\mathrm{M}_\odot$ (BHBs). In addition, we test the reliability of two methods using two mock galaxies from Auriga simulations, Au-6 and Au-16. Using the same procedure as applied to the real data, we find that the virial mass estimates deviate from the true mass of two Auriga galaxies by $15.5\%\pm2.8\%$ for the RRLs, and $11.9\%\pm3.4\%$ for the BHBs. After correcting for these systematic errors, the Milky Way virial masses are $M_{200} = 9.79^{+0.93}_{-0.84}\times10^{11}\,\mathrm{M}_\odot$ from RRLs and $M_{200} = 9.24^{+0.95}_{-0.84}\times10^{11}\,\mathrm{M}_\odot$ from BHBs.

In the coming years, large sky surveys including SDSS-V, DESI DR2, 4MOST and \textit{Gaia} DR4 will provide significantly larger samples of Milky Way stars reaching to greater Galactocentric distances and with significantly improved precision, enabling tighter constraints on the mass distribution of the Milky Way.

\begin{acknowledgments}
Y.H. acknowledges support from National Key R\&D Programme of China grant Nos. 2025YFF0510603, the National Science Foundation of China (NSFC grant No. 12422303), and the Fundamental Research Funds for the Central Universities (grant Nos. 118900M122, E5EQ3301X2, and E4EQ3301X2).

\textit{Software:} NumPy\citep{NumPy}, Matplotlib\citep{Matplotlib}, SciPy\citep{SciPy}, Astropy\citep{Astropy}, TOPCAT\citep{TOPCAT}, astroML\citep{astroML}

\end{acknowledgments}

\bibliography{paper}{} 
\bibliographystyle{aasjournal}

\end{document}